\documentclass[aps,prd,nofootinbib,showpacs,11pt]{revtex4}
\usepackage{txfonts}
\usepackage{graphicx}
\usepackage{dcolumn}
\usepackage{bm}
\usepackage{amssymb}
\usepackage{latexsym}

\newcommand{\be}{\begin{equation}}
\newcommand{\ee}{\end{equation}}
\newcommand{\bq}{\begin{eqnarray}}
\newcommand{\eq}{\end{eqnarray}}

\begin{document}

\title{Holographic Ricci dark energy: Current observational constraints, quintom feature,
and the reconstruction of scalar-field dark energy}

\author{Xin Zhang}
\affiliation{Department of Physics, College of Sciences,
Northeastern University, Shenyang 110004, China} \affiliation{Kavli
Institute for Theoretical Physics China, Chinese Academy of
Sciences, Beijing 100080, China}

\begin{abstract}

In this work, we consider the cosmological constraints on the
holographic Ricci dark energy proposed by Gao {\it et~al.} [Phys.
Rev. D 79, 043511 (2009)], by using the observational data currently
available. The main characteristic of holographic Ricci dark energy
is governed by a positive numerical parameter $\alpha$ in the model.
When $\alpha<1/2$, the holographic Ricci dark energy will exhibit a
quintomlike behavior; i.e., its equation of state will evolve across
the cosmological-constant boundary $w=-1$. The parameter $\alpha$
can be determined only by observations. Thus, in order to
characterize the evolving feature of dark energy and to predict the
fate of the universe, it is of extraordinary importance to constrain
the parameter $\alpha$ by using the observational data. In this
paper, we derive constraints on the holographic Ricci dark energy
model from the latest observational data including the Union sample
of 307 type Ia supernovae, the shift parameter of the cosmic
microwave background given by the five-year Wilkinson Microwave
Anisotropy Probe observations, and the baryon acoustic oscillation
measurement from the Sloan Digital Sky Survey. The joint analysis
gives the best-fit results (with 1$\sigma$ uncertainty):
$\alpha=0.359^{+0.024}_{-0.025}$ and $\Omega_{\rm
m0}=0.318^{+0.026}_{-0.024}$. That is to say, according to the
observations, the holographic Ricci dark energy takes on the quintom
feature. Finally, in light of the results of the cosmological
constraints, we discuss the issue of the scalar-field dark energy
reconstruction, based on the scenario of the holographic Ricci
vacuum energy.

\end{abstract}


\maketitle

\section{Introduction}\label{sec:intro}

The astronomical observations over the past decade imply that our
universe is currently dominated by dark energy that leads to an
accelerated expansion of the universe (see, e.g., Refs.
\cite{sn98,sdss04,wmap03}). The combined analysis of cosmological
observations suggests that the universe is spatially flat and
consists of about $70\%$ dark energy, $30\%$ dust matter (cold dark
matter plus baryons), and negligible radiation. Although we can
affirm that the ultimate fate of the universe is determined by the
feature of dark energy, the nature of dark energy as well as its
cosmological origin remain enigmatic at present (for reviews see,
e.g., \cite{DErev}). However, we still can propose some candidates
to interpret or describe the properties of dark energy. The most
obvious theoretical candidate of dark energy is the cosmological
constant $\lambda$ \cite{Einstein17}, which always suffers from the
``fine-tuning'' and ``cosmic coincidence'' puzzles. The fine-tuning
problem asks why the vacuum energy density today is so small
compared to typical particle scales. The vacuum energy density is of
order $10^{-47} {\rm GeV}^4$, which appears to require the
introduction of a new mass scale 14 or so orders of magnitude
smaller than the electroweak scale. The second difficulty, the
cosmic coincidence problem, says: Since the energy densities of
vacuum energy and dark matter scale so differently during the
expansion history of the universe, why are they nearly equal today?
To get this coincidence, it appears that their ratio must be set to
a specific, infinitesimal value in the very early universe.
Theorists have made lots of efforts to try to resolve the
cosmological-constant problem, but all of these efforts turned out
to be unsuccessful.

Numerous other candidates for dark energy have also been proposed in
the literature, such as an evolving canonical scalar field
\cite{quintessence} usually referred to as quintessence, the phantom
energy \cite{phantom} with an equation of state smaller than $-1$
violating the weak energy condition, the quintom energy
\cite{quintom,quintomext} with an equation of state evolving across
$-1$, and so forth.

Actually, the cosmological-constant (or dark energy) problem is in
essence an issue of quantum gravity because the cosmological
constant (or the dark energy density) is inevitably related to the
vacuum expectation value of some quantum fields within the
cosmological context. Therefore, in principle, we cannot entirely
understand the nature of dark energy before a complete theory of
quantum gravity is established. However, although we are lacking a
quantum gravity theory today, we still can make some attempts to
probe the nature of dark energy according to some principles of
quantum gravity. By far, the holographic principle is widely
believed as a fundamental principle for the theory of quantum
gravity that is being established. Hence, it is believed that the
holographic principle may play a significant role in shedding light
on the nature of the cosmological constant/ dark energy.

Currently, an interesting attempt for probing the nature of dark
energy within the framework of quantum gravity is the so-called
``holographic dark energy'' proposal \cite{Cohen99,Hsu04,Li04}. It
is well known that the holographic principle is an important result
of the recent research for exploring the quantum gravity (or string
theory) \cite{holop}. This principle is enlightened by
investigations of the quantum property of black holes. In a quantum
gravity system, the conventional local quantum field theory will
break down because it contains too many degrees of freedom that will
lead to the formation of a black hole breaking the effectiveness of
the quantum field theory. To reconcile this breakdown with the
success of local quantum field theory in describing observed
particle phenomenology, some authors proposed a relationship between
the ultraviolet (UV) and the infrared (IR) cutoffs due to the limit
set by the formation of a black hole. The UV-IR relation in turn
provides an upper bound on the zero-point energy density. In other
words, if the quantum zero-point energy density $\rho_{\rm vac}$ is
relevant to an UV cutoff, the total energy of the whole system with
size $L$ should not exceed the mass of a black hole of the same
size, and thus we have $L^3\rho_{\rm vac}\leq LM_{\rm Pl}^2$. This
means that the maximum entropy is of the order of $S_{BH}^{3/4}$.
When we take the whole universe into account, the vacuum energy
related to this holographic principle \cite{holop} is viewed as dark
energy, usually dubbed holographic dark energy (its density is
denoted as $\rho_{\rm de}$ hereafter). The largest IR cutoff $L$ is
chosen by saturating the inequality so that we get the holographic
dark energy density \cite{Li04}
\begin{equation}
\rho_{\rm de}=3c^2 M_{\rm Pl}^2L^{-2}~,\label{de}
\end{equation} where $c$ is a numerical constant, and
$M_{\rm Pl}\equiv 1/\sqrt{8\pi G}$ is the reduced Planck mass. If we
take $L$ as the size of the current universe, for instance the
Hubble radius $H^{-1}$, then the dark energy density will be close
to the observational result.

However, if one takes the Hubble scale as the IR cutoff, the
holographic dark energy seems not to be cable of leading to an
accelerating universe. The possibilities of the particle and the
event horizons as the IR cutoff were subsequently discussed by Li
\cite{Li04}, and it was found that only the event horizon acting as
the IR cutoff can give a viable holographic dark energy leading to
an accelerating universe. The holographic dark energy model based on
the event horizon as the IR cutoff has been widely studied
\cite{holoext,intholo} and found to be consistent with the
observational data \cite{holofitzx,holofitext}.

Although the holographic model based on the event horizon is
successful in fitting the current observational data, the model is
suffering from some serious conceptual problems. As discussed in
Ref.~\cite{Cai:2007us}, the event horizon may lead to an obvious
drawback concerning the causality. Since the event horizon is a
global concept of space-time, and the density of dark energy is,
however, a local quantity, a question naturally arises: Why should a
local quantity be determined by a global one? In addition, the event
horizon is determined by the future evolution of the universe,
leading to a puzzle of why the current density of dark energy is
determined by the future evolution of the universe rather than the
past of the universe. Moreover, the future event horizon can exist
only under the condition that the future evolution of the universe
is always in an acceleration phase, and thus it appears that a
causality problem is encountered, posting a challenge to the model.

To avoid the causality problem, it was proposed in
Ref.~\cite{Cai:2007us} that the age of the universe can be chosen as
the length measure, instead of the horizon distance of the universe.
In this case, by combining the general relativity and the
uncertainty relation in quantum mechanics, the energy density of
quantum fluctuations of space-time can be viewed as dark energy, and
this model is consistent with the observational data provided that
the unique parameter is taken to be a number of order unity. A new
version of this model replacing the age of the universe by the
conformal time of the universe was also discussed in
Ref.~\cite{newage}, in order to avoid some internal inconsistencies
in the original model. For further studies on this model, see, e.g.,
\cite{ageext}.

Furthermore, inspired by the above ideas on the holographic dark
energy, Gao {\it et~al.} \cite{Gao:2007ep} proposed to consider
another interesting possibility: The length scale, namely, the IR
cutoff, in the holographic model may be given by the average radius
of the Ricci scalar curvature ${|\cal R|}^{-1/2}$, so in this case
the density of the holographic dark energy is $\rho_{\rm de}\propto
{\cal R}$. This is the so-called ``holographic Ricci dark energy''
model. See also, e.g., \cite{ricciext}, for extensive studies. The
studies on its phenomenological properties show that this model
works fairly well in explaining the observations such as the cosmic
acceleration, and it could also help to understand the cosmic
coincidence problem. The model is free of the causality problem and
the age problem that plague the holographic model based on the
future event horizon. However, it should be pointed out that the
physical motivation for the Ricci model is still obscure in
Ref.~\cite{Gao:2007ep}.

Recently, however, Cai, Hu, and Zhang \cite{Cai:2008nk} investigated
the causal entropy bound in the holographic framework, providing us
with an appropriate physical motivation for the holographic Ricci
dark energy. The causal entropy bound for a spatial region in a
cosmological setting is given by assuming that the maximal black
hole in the universe is formed by gravitational collapse with the
``Jeans'' scale of perturbations, beyond which the black hole cannot
form very likely. Therefore, the Jeans scale of perturbations in the
universe gives a causal connection scale $R_{\rm CC}$ that is
naturally to be chosen as an IR cutoff in the holographic setup. For
gravitational perturbations, $R_{\rm CC}^{-2}={\rm
Max}(\dot{H}+2H^2,~-H)$ for a flat universe. It turns out that only
the case with the choice $R_{\rm CC}^{-2}=\dot{H}+2H^2$
(proportional to the Ricci scalar $\cal R$ of the
Friedmann-Robertson-Walker space-time in this case), could be
consistent with the current cosmological observations.

Like the Li model of the holographic dark energy, the main
characteristic of the holographic Ricci dark energy is also governed
by the numerical parameter $c$ in the model. In particular, when
$c^2<1/2$, the holographic Ricci dark energy will exhibit a
quintomlike behavior; i.e., its equation of state will evolve across
the cosmological-constant boundary $w=-1$. The parameter $c$ can be
determined only by observations. Thus, in order to characterize the
evolving feature of dark energy and to predict the fate of the
universe, it is of extraordinary importance to constrain the
parameter $c$ by using the observational data. Note that hereafter
we will use the redefined parameter $\alpha$ with $\alpha=c^2$ as in
Ref.~\cite{Gao:2007ep}. In this paper, we will use the observational
data currently available to constrain the parameters in the model of
holographic Ricci dark energy.

On the other hand, the scalar-field dark energy models are often
viewed as an effective description of the underlying theory of dark
energy. However, the underlying theory of dark energy cannot be
achieved before a complete theory of quantum gravity is established.
We can, nevertheless, speculate on the underlying theory of dark
energy by taking some principles of quantum gravity into account.
The holographic models of dark energy are no doubt tentative in this
way. We are now interested in, if we assume the holographic Ricci
vacuum energy scenario as the underlying theory of dark energy, how
the scalar-field model can be used to effectively describe it. We
will also address this issue in light of the fitting results to the
observational data.

This paper is organized as follows: In Sec.~\ref{sec:rde}, we review
the model of holographic Ricci dark energy and discuss the basic
characteristics of the model. In Sec.~\ref{sec:obs}, we perform
constraints on the holographic Ricci dark energy model by using the
up-to-date observational data sets. In Sec.~\ref{sec:reconstruct},
we discuss the issue of the reconstruction of the scalar-field dark
energy model from the observations, according to the scenario of the
holographic Ricci vacuum energy. Finally, we give the concluding
remarks in Sec.~\ref{sec:concl}.

\section{The model of holographic Ricci dark energy}\label{sec:rde}

In this section, we briefly review the model of the holographic
Ricci dark energy. We first consider the general case with an
arbitrary spatial geometry in the Friedmann-Robertson-Walker (FRW)
universe, and then in practice we focus only on the spatially flat
case as motivated by the inflation.

Consider the FRW universe with the metric
\begin{equation}
ds^2=-dt^2+a(t)^2\left({dr^2\over 1-kr^2}+r^2
d\theta^2+r^2\sin^2\theta d\phi^2\right),
\end{equation}
where $k=1$, 0, $-1$ for closed, flat, and open geometries,
respectively, and $a(t)$ is the scale factor of the universe with
the convention $a(t_0)=1$. The Friedmann equation is
\begin{equation}
H^2={8\pi G\over 3}\sum\limits_i \rho_i -{k\over a^2},
\end{equation}
where $H=\dot{a}/a$ is the Hubble parameter, the dot denotes the
derivative with respect to the cosmic time $t$, and the summation
runs over various cosmic components. If we focus only on the
late-time evolution of the universe, the radiation component
$\rho_{\rm rad}$ is negligible, and then the cosmic contents include
the matter component $\rho_{\rm m}$ and the dark energy component
$\rho_{\rm de}$. The Ricci scalar
\begin{equation}
{\cal R}=-6\left(\dot{H}+2H^2+{k\over a^2}\right)
\end{equation}
determines, as suggested by Gao {\it et~al.} \cite{Gao:2007ep}, the
density of dark energy:
\begin{equation}
\rho_{\rm de}={3\alpha\over 8\pi G}\left(\dot{H}+2H^2+{k\over
a^2}\right)=-{\alpha\over 16\pi G}{\cal R},
\end{equation}
where $\alpha$ is a positive numerical constant to be determined by
observations. Comparing to Eq.~(\ref{de}), we see that if we
identify $L^{-2}$ with $-{\cal R}/6$, we have $\alpha=c^2$. This is
the so-called holographic Ricci dark energy model. This model was
originally viewed as lacking physical reasoning \cite{Gao:2007ep}.
Thanks to the work of Cai {\it et~al.} \cite{Cai:2008nk}, the Ricci
model gets an appropriate physical mechanism or reasoning for which
such a dark energy could be motivated. Assuming the maximal black
hole in the universe is formed through gravitational collapse of
perturbations in the universe, then the Jeans scale of the
perturbations gives a causal connection scale $R_{\rm CC}$ that is
naturally to be chosen as an IR cutoff in the holographic setup. For
gravitational perturbations, $R_{\rm CC}^{-2}={\rm
Max}(\dot{H}+2H^2,~-H)$ for a flat universe. It turns out that only
the case with the choice $R_{\rm CC}^{-2}=\dot{H}+2H^2$
(proportional to the Ricci scalar $\cal R$ of the FRW space-time in
this case) could be consistent with the current cosmological
observations. So, the Ricci dark energy can be viewed as originating
from taking the causal connection scale as the IR cutoff in the
holographic setting.

Now the Friedmann equation can be rewritten as
\begin{equation}
H^2={8\pi G\over 3}\rho_{\rm m0}e^{-3x}+(\alpha-1)k
e^{-2x}+\alpha\left({1\over 2}{dH^2\over dx}+2H^2\right),
\end{equation}
where $x=\ln a$ and the subscript ``0'' denotes the present values
of various variables, hereafter. This equation can be further
rewritten in the following form:
\begin{equation}
E^2=\Omega_{\rm m0}e^{-3x} +(1-\alpha)\Omega_{k0}
e^{-2x}+\alpha\left({1\over 2}{dE^2\over dx}+2E^2\right),
\end{equation}
where $E\equiv H/H_0$, $\Omega_{\rm m0}=8\pi G\rho_{\rm m0}/(3H^2)$
and $\Omega_{k0}=-k/H_0^2$. Solving this equation, one obtains
\begin{equation}
E(a)^2=\Omega_{\rm m0}a^{-3}+\Omega_{k0}a^{-2}+{\alpha\over
2-\alpha}\Omega_{\rm m0}a^{-3}+f_0 a^{-(4-{2\over
\alpha})},\label{Ea}
\end{equation}
where $f_0$ is an integration constant. Using the initial condition
$E_0=E(t_0)=1$, the integration constant $f_0$ is determined as
\begin{equation}
f_0=1-\Omega_{k0}-{2\over 2-\alpha}\Omega_{\rm m0}.
\end{equation}

In Eq.~(\ref{Ea}), it is easy to identify the contribution of dark
energy (the last two terms on the right hand side); consequently, we
can define
\begin{equation}
\tilde{\Omega}_{\rm de}\equiv {\rho_{\rm de}\over \rho_{\rm
c0}}={\alpha\over 2-\alpha}\Omega_{\rm m0}a^{-3}+f_0 a^{-(4-{2\over
\alpha})},\label{rdet}
\end{equation}
where $\rho_{\rm c0}=3H_0^2/(8\pi G)$ is the present critical
density of the universe. From this expression, one can see that the
parameter $\alpha$ plays a significant role for the evolution of the
Ricci dark energy. When $1/2\leq\alpha\leq1$, the equation of state
of dark energy will evolve in the region of $-1\leq w\leq -1/3$. In
particular, if $\alpha=1/2$ is chosen, the behavior of the
holographic Ricci dark energy will be more and more like a
cosmological constant with the expansion of the universe, such that
ultimately the universe will enter the de Sitter phase in the far
future. When $\alpha<1/2$, the holographic Ricci dark energy will
exhibit a quintomlike evolution behavior (for ``quintom'' dark
energy, see, e.g., \cite{quintom} and references therein), i.e., the
equation of state of holographic Ricci dark energy will evolve
across the cosmological-constant boundary $w=-1$ (actually, it
evolves from the region with $w>-1$ to that with $w<-1$). That is to
say, the choice of $\alpha<1/2$ makes the Ricci dark energy today
behave as a phantom energy that leads to a cosmic doomsday (``big
rip'') in the future. Thus, as discussed above, the value of
$\alpha$ determines the destiny of the universe in the holographic
Ricci dark energy model. On the other hand, from Eq.~(\ref{rdet}),
one can easily infer that the Ricci dark energy could track the
evolution of the nonrelativistic matter in the early times, which
can help to alleviate the cosmic coincidence problem.

\begin{figure}[htbp]
\begin{center}
\includegraphics[scale=1.2]{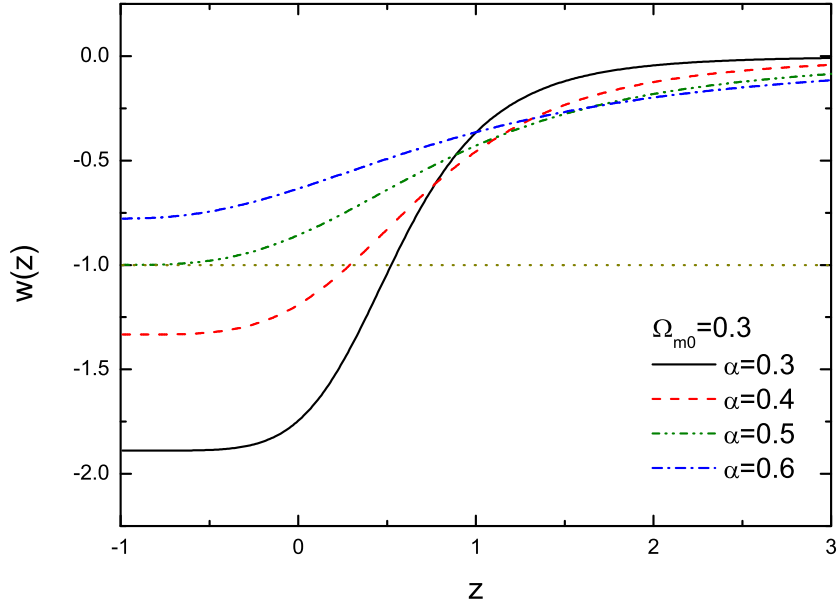}
\caption{The evolution of the equation of state parameter for the
holographic Ricci dark energy. Here we take $\Omega_{\rm m0}=0.3$
and show the cases for $\alpha=0.3$, 0.4, 0.5, and 0.6. Clearly, the
cases with $\alpha\geq 1/2$ behave like a quintessence, and the
cases with $\alpha<1/2$ behave like a quintom.}\label{fig:wz}
\end{center}
\end{figure}

\begin{figure}[htbp]
\begin{center}
\includegraphics[scale=1.2]{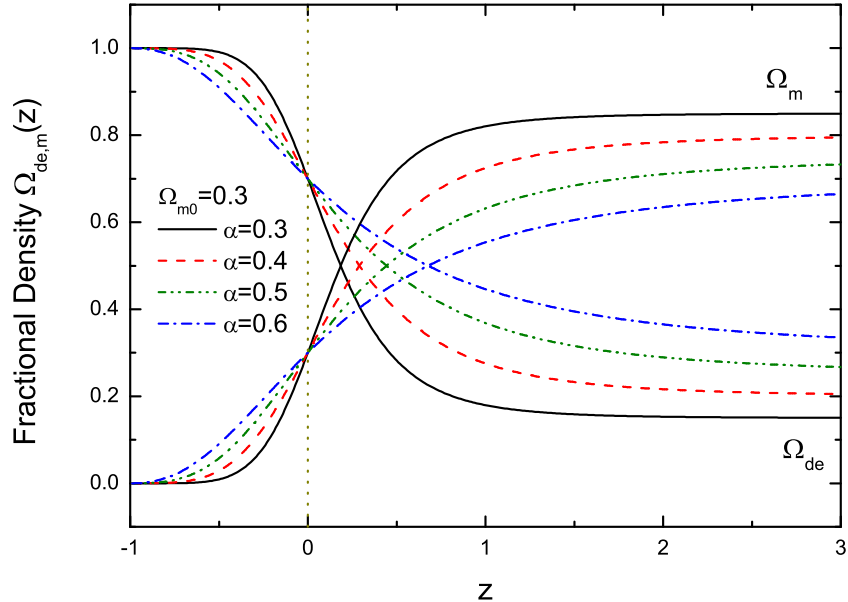}
\caption{The evolution of the fractional densities $\Omega_{\rm
de}(z)$ and $\Omega_{\rm m}(z)$. Also, we plot the cases for
$\Omega_{\rm m0}=0.3$ and $\alpha=0.3$, 0.4, 0.5, and 0.6. At first
sight, one finds that at early times of roughly $z>2$ the density
contribution of dark energy can occupy roughly $20\%-30\%$. However,
it should be pointed out that in this epoch the dark energy behaves
like dust matter, so, effectively speaking, the matter density
contribution is still $\sim 100\%$, namely, $\Omega_{\rm m}^{\rm
(eff)}\sim 1$. }\label{fig:Omegaz}
\end{center}
\end{figure}

Of course, one can also derive the usual fractional density of dark
energy,
\begin{equation}
\Omega_{\rm de}\equiv{\rho_{\rm de}\over \rho_{\rm
c}}={\tilde{\Omega}_{\rm de}\over E^2},
\end{equation}
where $\rho_{\rm c}=3H^2/(8\pi G)$ is the critical density of the
universe. Furthermore, from the energy conservation equation
$\dot{\rho}_{\rm de}+3H(1+w)\rho_{\rm de}=0$, one can obtain the
equation of state for Ricci dark energy
\begin{equation}
w(z)=-1+{(1+z)\over 3}{d\ln \tilde{\Omega}_{\rm de}\over dz},
\end{equation}
where $z=(1/a)-1$ is the redshift.

Since the current observations strongly favor a spatially flat
universe that is also supported by the inflation theory, hereafter
the discussions will be restricted to the case of $\Omega_{k0}=0$
(or $k=0$).

As illustrative examples, we plot in Figs.~\ref{fig:wz} and
\ref{fig:Omegaz} the selected evolutions of the holographic Ricci
dark energy. Figure~\ref{fig:wz} shows the evolution of the equation
of state $w(z)$, and Fig.~\ref{fig:Omegaz} shows the evolution of
the fractional densities $\Omega_{\rm de}(z)$ and $\Omega_{\rm
m}(z)$. In both figures, we plot the cases for $\Omega_{\rm m0}=0.3$
and $\alpha=0.3$, 0.4, 0.5 and 0.6. From Fig. \ref{fig:wz}, it is
clear that the cases with $\alpha\geq 1/2$ always evolve in the
region of $w\geq -1$, whereas the cases with $\alpha<1/2$ behave as
a quintom whose equation of state $w$ crosses the
cosmological-constant boundary $-1$ during the evolution. Also, at
early times, roughly $z>2$, the equation of state of the Ricci dark
energy approaches 0; i.e., in this model the dark energy behaves
like dust matter during most of the epoch of matter domination. This
tracking behavior can help to alleviate the cosmic coincidence
problem of dark energy. In Fig.~\ref{fig:Omegaz}, one finds that at
early times of roughly $z>2$ the density contribution of dark energy
can occupy roughly $20\%-30\%$. However, it should be pointed out
that in this epoch the dark energy behaves like dust matter, so,
effectively speaking, the matter density contribution is still $\sim
100\%$, namely, $\Omega_{\rm m}^{\rm (eff)}\sim 1$, almost the same
as the $\Lambda$CDM model. It has been shown in
Ref.~\cite{Gao:2007ep} that the structure formation in this model is
very similar to that in the $\Lambda$CDM model.

\section{Current observational constraints}\label{sec:obs}

In this section, we constrain the parameters in the holographic
Ricci dark energy model and analyze the evolutionary behavior of
this dark energy by using the latest observational data of type Ia
supernova (SNIa) combined with the information from cosmic microwave
background (CMB) and large scale structure (LSS) observations.

\subsection{Cosmological constraints from SNIa}

First, we consider the latest 307 Union SNIa data, the distance
modulus $\mu_{\rm obs}(z_i)$, compiled in \cite{sn08}. The SCP
(Supernova Cosmology Project) ``Union'' SNIa compilation brings
together data from 414 SNe drawn from 13 independent data sets, of
which 307 SNe pass usability cuts. All SNe were fit using a single
lightcurve fitter and uniformly analyzed. All analyses and cuts were
developed in a blind manner, i.e. with the cosmology hidden. We
shall analyze the holographic Ricci dark energy model in light of
the Union sample of SNIa in this subsection.

The theoretical distance modulus is defined as
\begin{equation}
\mu_{\rm th}(z_i)\equiv 5 \log_{10} {D_L(z_i)} +\mu_0,
\end{equation}
where $\mu_0\equiv 42.38-5\log_{10}h$, $h$ is the Hubble constant
$H_0$ in units of 100 km/s/Mpc, and
\begin{equation}
D_L(z)=(1+z)\int_0^z {dz'\over E(z';{\bm \theta})}
\end{equation}
is the Hubble-free luminosity distance $H_0d_L$ (here $d_L$ is the
physical luminosity distance) in a spatially flat FRW universe,
where ${\bm \theta}$ denotes the model parameters.

The $\chi^2$ for the SNIa data is
\begin{equation}
\chi^2_{\rm SN}({\bm\theta})=\sum\limits_{i=1}^{307}{[\mu_{\rm
obs}(z_i)-\mu_{\rm th}(z_i)]^2\over \sigma_i^2},\label{ochisn}
\end{equation}
where $\mu_{\rm obs}(z_i)$ and $\sigma_i$ are the observed value and
the corresponding 1$\sigma$ error for each supernova, respectively.
The parameter $\mu_0$ is a nuisance parameter, but it is independent
of the data points and the data set. Following
\cite{Nesseris:2005ur}, the minimization with respect to $\mu_0$ can
be made trivially by expanding the $\chi^2$ of Eq.~(\ref{ochisn})
with respect to $\mu_0$ as
\begin{equation}
\chi^2_{\rm SN}({\bm\theta})=A({\bm\theta})-2\mu_0
B({\bm\theta})+\mu_0^2 C,
\end{equation}
where
\begin{equation}
A({\bm\theta})=\sum\limits_{i=1}^{307}{[\mu_{\rm obs}(z_i)-\mu_{\rm
th}(z_i;\mu_0=0,{\bm\theta})]^2\over \sigma_i^2},
\end{equation}
\begin{equation}
B({\bm\theta})=\sum\limits_{i=1}^{307}{\mu_{\rm obs}(z_i)-\mu_{\rm
th}(z_i;\mu_0=0,{\bm\theta})\over \sigma_i^2},
\end{equation}
\begin{equation}
C=\sum\limits_{i=1}^{307}{1\over \sigma_i^2}.
\end{equation}
Evidently, Eq.~(\ref{ochisn}) has a minimum for $\mu_0=B/C$ at
\begin{equation}
\tilde{\chi}^2_{\rm
SN}({\bm\theta})=A({\bm\theta})-{B({\bm\theta})^2\over
C}.\label{tchi2sn}
\end{equation}
Since $\chi^2_{\rm SN, min}=\tilde{\chi}^2_{\rm SN, min}$, instead
minimizing $\chi^2_{\rm SN}$ one can minimize $\tilde{\chi}^2_{\rm
SN}$ which is independent of the nuisance parameter $\mu_0$.
Obviously, the best-fit value of $h$ can be given by the
corresponding $\mu_0=B/C$ at the best fit.

\begin{figure}[htbp]
\begin{center}
\includegraphics[scale=1.2]{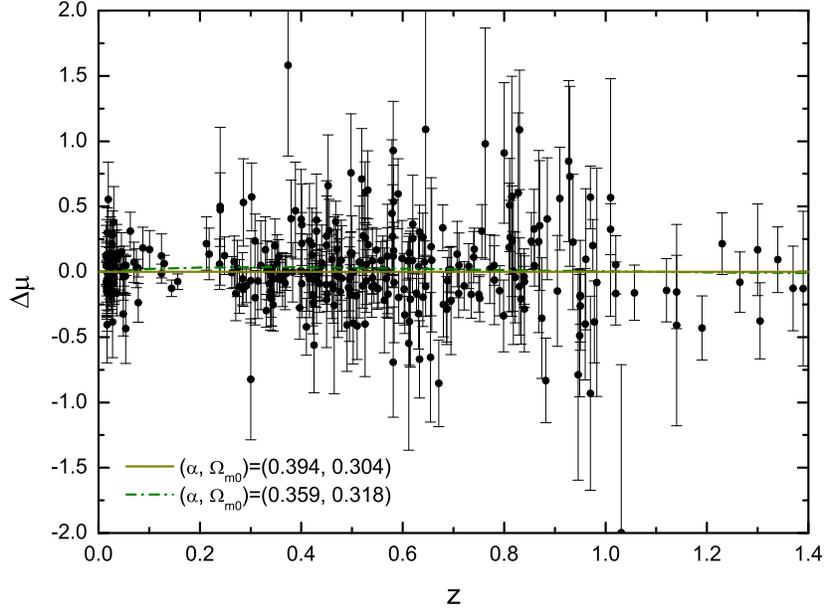}
\caption{SCP Union sample of 307 SNIa residual Hubble diagram
comparing to the holographic Ricci dark energy model with best-fit
values for parameters. The dark-yellow solid line represents the
best fit for SNIa alone analysis with $(\alpha,~\Omega_{\rm
m0})=(0.394,~0.304)$; the green dashed line represents the best-fit
for SNIa+CMB+BAO joint analysis with $(\alpha,~\Omega_{\rm
m0})=(0.359,~0.318)$. The data and model are shown relative to the
case of $(\alpha,~\Omega_{\rm
m0})=(0.394,~0.304)$.}\label{fig:hubble}
\end{center}
\end{figure}

\begin{figure}[htbp]
\begin{center}
\includegraphics[scale=0.5]{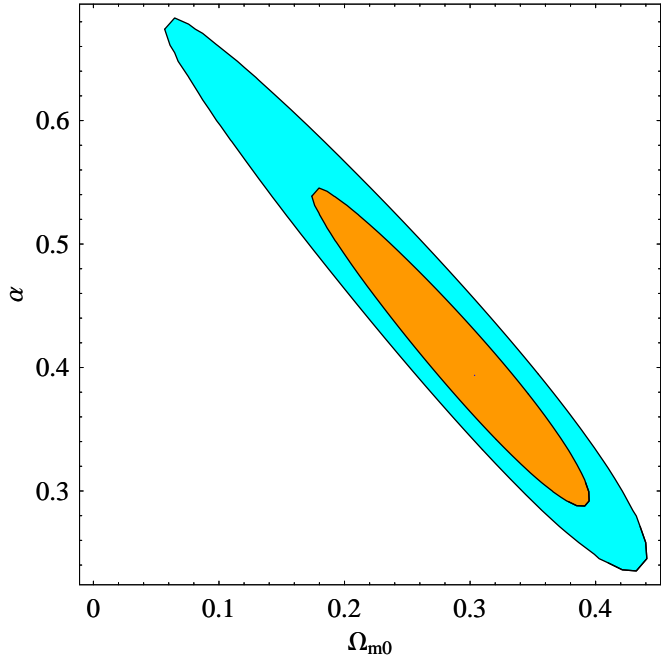}
\caption{Probability contours at $68.3\%$ and $95.4\%$ confidence
levels in the $(\Omega_{\rm m0},~\alpha)$ plane, for the holographic
Ricci dark energy model, from the constraints of the SCP Union SNIa
data. The fit values for model parameters with one-sigma errors are
$\alpha=0.394^{+0.152}_{-0.106}$ and $\Omega_{\rm
m0}=0.304^{+0.091}_{-0.131}$. A point denotes the best fit; at the
best fit, we have $\chi^2_{\rm min}=310.682$ and
$h=0.704$.}\label{fig:sncontour}
\end{center}
\end{figure}

\begin{figure}[htbp]
\begin{center}
\includegraphics[scale=0.5]{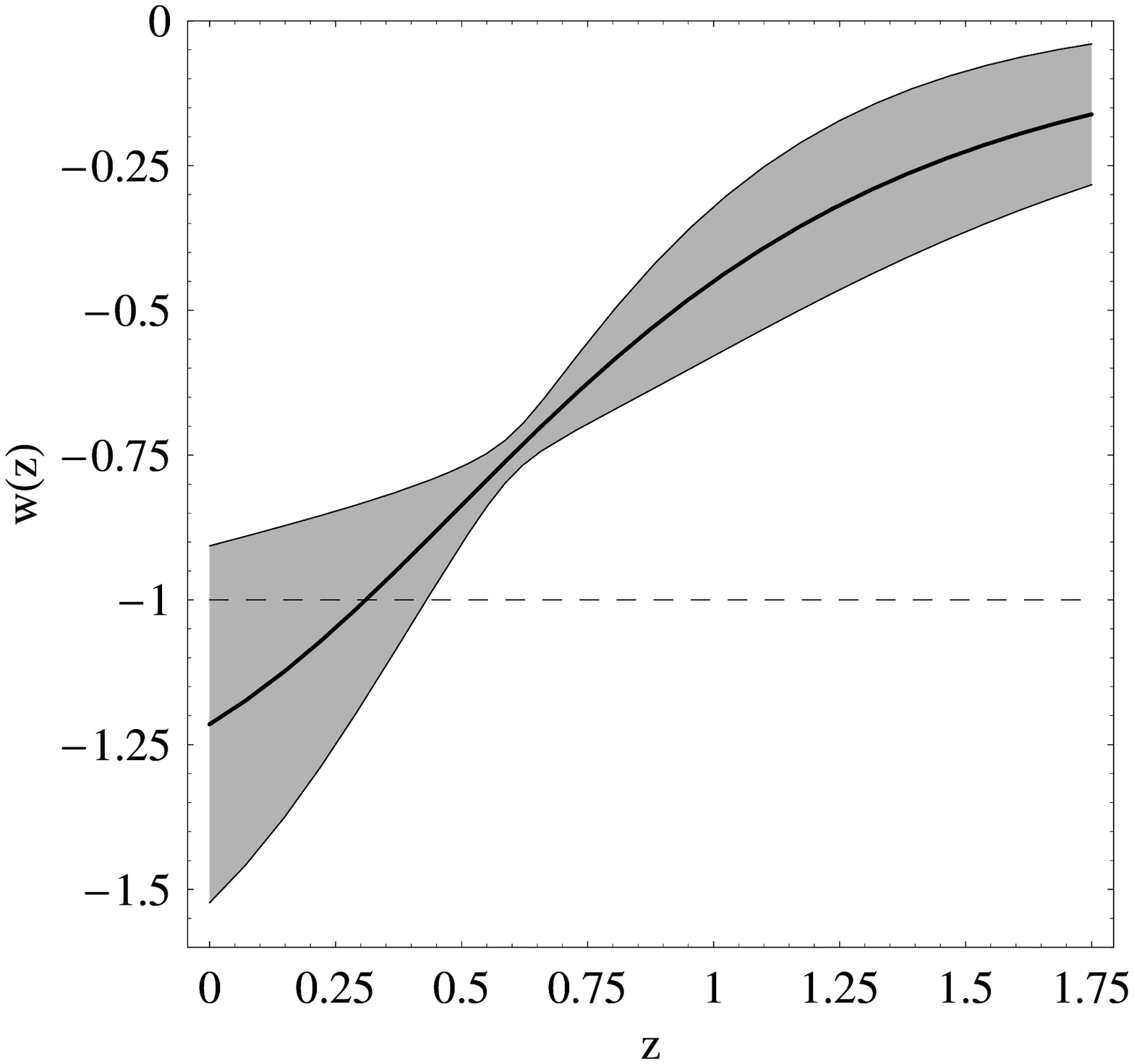}
\caption{Constraints on the evolution of the equation of state
$w(z)$ from the SNIa data. The central thick solid line represents
the best fit, and the light gray contour represents the 1$\sigma$
confidence level around the best fit. The present value of the
equation of state $w_0$, with 1$\sigma$ error, is $w_0=-1.215\pm
0.308$. Errors are calculated by the Fisher matrix
approach.}\label{fig:wzsn}
\end{center}
\end{figure}

The best fit for the analysis of the SCP Union sample of 307 SNIa
happens at $\alpha=0.394$, $\Omega_{\rm m0}=0.304$, and $h=0.704$,
with $\chi^2_{\rm min}=310.682$. The Union sample is illustrated on
a residual Hubble diagram with respect to our best-fit universe in
Fig.~\ref{fig:hubble}. Next, we show the probability contours at
$68.3\%$ and $95.4\%$ confidence levels for $\alpha$ versus
$\Omega_{\rm m0}$ in Fig.~\ref{fig:sncontour}, from the constraints
of the SNIa data. The 1$\sigma$ and 2$\sigma$ fit values for the
model parameters are
$\alpha=0.394^{+0.152}_{-0.106}~(1\sigma)^{+0.290}_{-0.159}~(2\sigma)$
and $\Omega_{\rm
m0}=0.304^{+0.091}_{-0.131}~(1\sigma)^{+0.137}_{-0.248}~(2\sigma)$.
We see that the best-fit value for parameter $\alpha$ is 0.394,
smaller than 0.5, leading the holographic Ricci dark energy to
behave as a quintom with equation of state evolving across $w=-1$.
Moreover, the parameter $\alpha$ in 1$\sigma$ range,
$0.288<\alpha<0.546$, is also basically smaller than 0.5, albeit the
1$\sigma$ upper bound slightly larger than 0.5, indicating the
quintom nature for the holographic Ricci dark energy. To see the
constraints on the evolution of the equation of state from the SNIa
data, we show in Fig.~\ref{fig:wzsn} the corresponding $w(z)$ with
1$\sigma$ uncertainty. The present value of the equation of state
$w_0$, with 1$\sigma$ error, is $w_0=-1.215\pm 0.308$.

From Figs.~\ref{fig:sncontour} and \ref{fig:wzsn}, we see that the
SNIa data alone do not seem to be sufficient to constrain the
holographic Ricci dark energy model strictly. The confidence region
of the $\Omega_{\rm m0}-\alpha$ plane is rather large; say, the
2$\sigma$ ranges for the parameters are $\alpha\in (0.235,~0.684)$
and $\Omega_{\rm m0}\in (0.056,~0.441)$. To break the degeneracy of
the parameters, we seek to find other observations as complements to
the SNIa data. So, in the next subsection, we shall make a combined
analysis of SNIa, CMB, and LSS for the model of holographic Ricci
dark energy.

\subsection{Cosmological constraints from SNIa, CMB, and BAO}

In this subsection, we further perform constraints on the model of
holographic Ricci dark energy by combining the observations from
SNIa, CMB and LSS. For the CMB data, we use the CMB shift parameter
$R$; for the LSS data, we use the parameter $A$ of the baryon
acoustic oscillation (BAO) measurement. In fact, it is commonly
believed that both $R$ and $A$ are nearly model-independent and
contain essential information of the full CMB and LSS BAO data
(however, see also, e.g.,
\cite{Elgaroy:2007bv,baopeak,Rydbeck:2007gy}).

The shift parameter $R$ is given by \cite{Bond97,Wangyun:2006ts}
\begin{equation}
R\equiv \Omega_{\rm m0}^{1/2}\int_0^{z_{\rm CMB}}{dz'\over E(z')},
\end{equation}
where the redshift of recombination $z_{\rm CMB}=1090$ has been
updated in the Wilkinson Microwave Anisotropy Probe (WMAP) five-year
data \cite{Komatsu:2008hk}. The shift parameter $R$ relates the
angular diameter distance to the last scattering surface, the
comoving size of the sound horizon at $z_{\rm CMB}$, and the angular
scale of the first acoustic peak in the CMB power spectrum of
temperature fluctuations \cite{Bond97,Wangyun:2006ts}. The value of
the shift parameter $R$ has been updated by WMAP5
\cite{Komatsu:2008hk} to be $1.710\pm 0.019$ independent of dark
energy model. The parameter $A$ of the measurement of the BAO peak
in the distribution of Sloan Digital Sky Survey (SDSS) luminous red
galaxies is defined as
\begin{equation}
A\equiv \Omega_{\rm m0}^{1/2} E(z_{\rm BAO})^{-1/3}\left[{1\over
z_{\rm BAO}}\int_0^{z_{\rm BAO}}{dz'\over E(z')}\right]^{2/3},
\end{equation}
where $z_{\rm BAO}=0.35$. The SDSS BAO measurement \cite{bao05}
gives $A=0.469(n_s/0.98)^{-0.35}\pm 0.017$ (independent of a dark
energy model), where the scalar spectral index is taken to be
$n_s=0.960$ as measured by WMAP5 \cite{Komatsu:2008hk}. We notice
that both $R$ and $A$ are independent of $H_0$; thus these
quantities can provide a robust constraint as a complement to SNIa
data on the holographic Ricci dark energy model.

\begin{figure}[htbp]
\begin{center}
\includegraphics[scale=0.5]{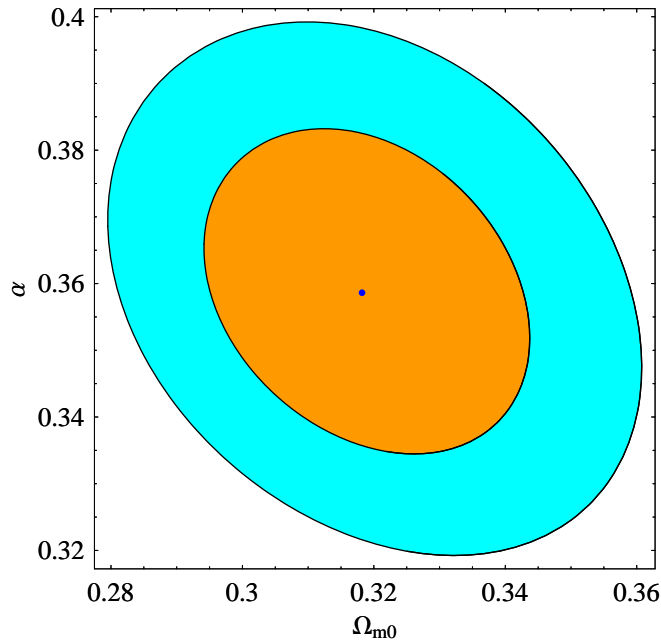}
\caption{Probability contours at $68.3\%$ and $95.4\%$ confidence
levels in the $(\Omega_{\rm m0},~\alpha)$ plane, for the holographic
Ricci dark energy model, from the joint analysis of the SNIa, CMB,
and BAO observations. The fit values for model parameters with
one-sigma errors are $\alpha=0.359^{+0.024}_{-0.025}$ and
$\Omega_{\rm m0}=0.318^{+0.026}_{-0.024}$. A point denotes the best
fit; at the best fit, we have $\chi^2_{\rm min}=324.317$ and
$h=0.711$.}\label{fig:contour}
\end{center}
\end{figure}

\begin{figure}[htbp]
\begin{center}
\includegraphics[scale=0.5]{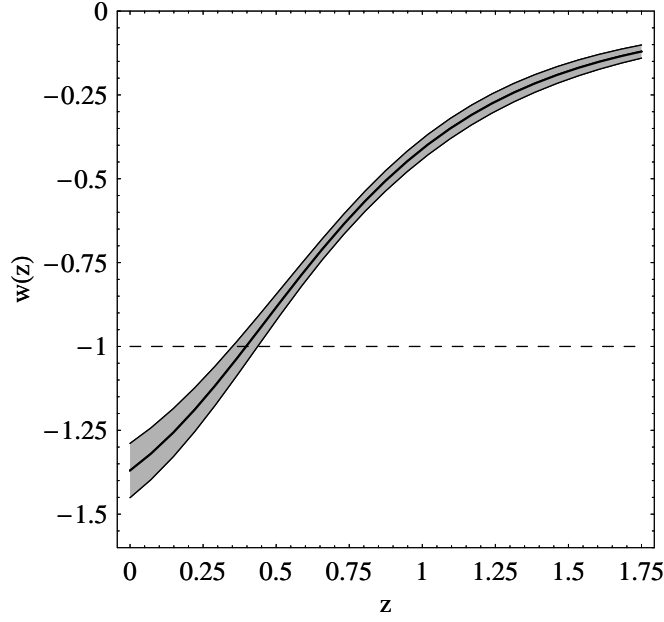}
\caption{Constraints on the evolution of the equation of state
$w(z)$, from the joint analysis of the SNIa, CMB, and BAO
observations. The central thick solid line represents the best fit,
and the light gray contour represents the 1$\sigma$ confidence level
around the best fit. The present value of the equation of state
$w_0$, with 1$\sigma$ error, is $w_0=-1.370\pm 0.081$. Errors are
calculated by the Fisher matrix approach. The quintom feature with
$w=-1$ crossing characteristic for the holographic Ricci dark energy
can be explicitly seen in this plot. }\label{fig:wzerr}
\end{center}
\end{figure}

Here, we pause for a while to make some additional comments on the
utilization of the SDSS baryon acoustic peak. Actually, about
whether or not the baryon acoustic peak should be used to constrain
models of dark energy that behave differently to a cosmological
constant, there is still some debate \cite{baopeak,Rydbeck:2007gy}.
The reason comes from the assumption of a constant equation of state
made in the reconstruction from redshift space to comoving space
required to accurately identify the position of the acoustic
peak~\cite{bao05}. For an alternative dark energy model where the
equation of state is a function of redshift, actually, it would be
expected that the change in the position of the acoustic peak is
small \cite{Rydbeck:2007gy}. Although so, it is indeed difficult to
quantify the correction without detailed study for each model in
question. However, it should also be pointed out that the SDSS
baryon acoustic peak has been adopted by the majority of the
cosmology community in placing constraint on dark energy models.
Therefore, in this paper, we do use the parameter $A$ of the BAO to
constrain the parameter space of the holographic Ricci dark energy
model, believing that it is indeed nearly model-independent.

We now perform a joint analysis of SNIa, CMB, and BAO on the
constraints of the holographic Ricci dark energy model. The total
$\chi^2$ is given by
\begin{equation}
\chi^2=\tilde{\chi}_{\rm SN}^2+\chi_{\rm CMB}^2+\chi_{\rm BAO}^2~,
\end{equation}
where $\tilde{\chi}_{\rm SN}^2$ is given by Eq.~(\ref{tchi2sn}) for
SNIa statistics and $\chi_{\rm CMB}^2=[({R}-{R}_{\rm
obs})/\sigma_{R}]^2$ and $\chi_{\rm BAO}^2=[(A-A_{\rm
obs})/\sigma_A]^2$ are contributions from CMB and BAO, respectively.

The main fitting result is shown in Fig.~\ref{fig:contour}. In this
figure, we show the contours of $68.3\%$ and $95.4\%$ confidence
levels in the $\Omega_{\rm m0}-\alpha$ plane. It is clear to see
that the combined analysis of SNIa, CMB, and BAO data provides a
fairly tight constraint on the holographic Ricci dark energy model,
compared to the constraint from the SNIa data alone. The fit values
for the model parameters with 1- and 2-$\sigma$ errors are
$\alpha=0.359^{+0.024}_{-0.025}~(1\sigma)^{+0.040}_{-0.040}~(2\sigma)$
and $\Omega_{\rm
m0}=0.318^{+0.026}_{-0.024}~(1\sigma)^{+0.043}_{-0.038}~(2\sigma)$
with $\chi^2_{\rm min}=324.317$. At the best fit, we have $h=0.711$.
We also show the best-fit case of SNIa+CMB+BAO analysis on the
residual Hubble diagram with respect to the best-fit case of SNIa
alone analysis in Fig.~\ref{fig:hubble}. As a comparison, we also
fit the spatially flat $\Lambda$CDM model to the same observational
data. It is found that, for the $\Lambda$CDM model, we have
$\chi_{\rm min}^2=313.742$ for the best-fit parameter $\Omega_{\rm
m0}=0.270$.

From Fig.~\ref{fig:contour}, we see that, according to the joint
analysis of the observational data, the holographic Ricci dark
energy takes on the nature of a quintom, since the parameter
$\alpha$ is less than 0.5, say, in the 2$\sigma$ range, $\alpha\in
(0.319,~0.399)$. This result completely rules out the probability of
$\alpha>0.5$ and clarifies the ambiguity in the analysis of SNIa
alone. So, the joint analysis definitely concludes that the
holographic dark energy behaves as a quintom. The resulting $w(z)$
with 1$\sigma$ error is shown in Fig.~\ref{fig:wzerr}. In this
figure, the quintom feature with the $w=-1$ crossing characteristic
for the holographic Ricci dark energy can be explicitly seen. The
present value of $w_0$, with a 1$\sigma$ error, is $w_0=-1.370\pm
0.081$.

\section{The reconstruction of scalar-field dark
energy}\label{sec:reconstruct}

As explained by Cai {\it et~al.} \cite{Cai:2008nk}, the Ricci dark
energy takes the causal connection scale in the universe as the IR
cutoff in the holographic setting. When taking the holographic
principle into account, the vacuum energy will acquire a dynamical
property that its equation of state is evolving, as shown in the
previous sections. The current available observational data imply
that the holographic Ricci vacuum energy behaves as quintom-type
dark energy. Presently, we adopt the viewpoint that the scalar-field
models of dark energy are effective theories of an underlying theory
of dark energy. If we regard the scalar-field model as an effective
description of such a holographic vacuum theory, we should be
capable of using the scalar-field model to mimic the evolving
behavior of the dynamical vacuum energy and reconstructing this
scalar-field model according to the fits of the observational data
sets. In this section, we shall discuss this issue.

\subsection{Motivation for reconstruction}

It is well known that the cosmological-constant/dark energy problem
is an UV problem. However, when considering the holographic property
of gravity, the UV regime is related to the IR regime. Thanks to the
UV-IR relation, the dark energy problem can be converted to an IR
problem. This is the key point of the holographic dark energy
proposal. In this view, the UV-IR relation provides an upper bound
on the zero-point energy (vacuum energy) density, and consequently
the vacuum energy becomes a dynamical dark energy.

Actually, the dynamical dark energy scenario is an alternative
proposal to the cosmological-constant scenario. The dynamical dark
energy proposal is often realized by some scalar field mechanism
which suggests that the energy form with negative pressure is
provided by a scalar field evolving down a proper potential.
Actually, this mechanism is enlightened to a great extent by the
inflationary cosmology. As we have known, the occurrence of the
current accelerating expansion of the universe is not the first time
for the expansion history of the universe. There is significant
observational evidence strongly supporting that the universe
underwent an early inflationary epoch, over sufficiently small time
scales, during which its expansion rapidly accelerated under the
drive of an ``inflaton'' field which had properties similar to those
of a cosmological constant. The inflaton field, to some extent, can
be viewed as a kind of dynamically evolving dark energy. Hence, the
scalar-field models involving a minimally coupled scalar field are
proposed, inspired by inflationary cosmology, to construct
dynamically evolving models of dark energy. The only difference
between the dynamical scalar-field dark energy and the inflaton is
the energy scale that they possess. Famous examples of scalar-field
dark energy models include quintessence \cite{quintessence},
$K$-essence \cite{kessence}, tachyon \cite{tachyon}, phantom
\cite{phantom}, ghost condensate \cite{ghost} and two-field quintom
\cite{quintom}, and so forth.

Generically, there are two points of view on the scalar-field models
of dynamical dark energy. One viewpoint regards the scalar field as
a fundamental field of the nature. The nature of dark energy is,
according to this viewpoint, completely attributed to some
fundamental scalar field which is omnipresent in supersymmetric
field theories and in string/M theory. The other viewpoint supports
that the scalar-field model is an effective description of an
underlying theory of dark energy. On the whole, it seems that the
latter is the mainstream point of view. Since we regard the scalar
field model as an effective description of an underlying theory of
dark energy, a question arises: What is the underlying theory of the
dark energy? Of course, hitherto, this question is far beyond our
present knowledge, because we cannot entirely understand the nature
of dark energy before a complete theory of quantum gravity is
established.

Although we are lacking a quantum gravity theory today, we can,
nevertheless, speculate on the underlying theory of dark energy by
taking some principles of quantum gravity into account. Needless to
say, the holographic models of dark energy are an interesting
tentative in this way. Since the holographic principle is taken into
account, the holographic models possess some significant features of
an underlying theory of dark energy. Now, we are interested in, if
we assume the holographic Ricci dark energy as the underlying theory
of dark energy, how the scalar-field model can be used to describe
it. In Sec.~\ref{sec:obs}, we have constrained the holographic Ricci
dark energy model using the latest observational data. Hence, in
turn, if there is a low-energy effective scalar-field describing the
Ricci dark energy, the scalar-field model can be reconstructed in
light of the constraint results from the observations. For the works
in this way, see, e.g., \cite{holoreczx1,holoreczx2}.

\subsection{Reconstructing a single-scalar-field quintom model from the
observations}

The nomenclature quintom is suggested in the sense that its behavior
resembles the combined behavior of quintessence and phantom. Thus, a
simple realization of a quintom scenario is a model with the double
fields of quintessence and phantom \cite{quintom}. The cosmological
evolution of such a model has been investigated in detail. It should
be noted that such a quintom model would typically encounter the
problem of quantum instability inherited from the phantom component.

For the single real scalar-field models, the transition of crossing
$-1$ for $w$ can occur for the Lagrangian density $p(\phi, X)$,
where X is a kinematic term of a scalar-field $\phi$, in which
$\partial{p}/\partial{X}$ changes sign from positive to negative,
and thus we require nonlinear terms in $X$ to realize the $w=-1$
crossing \cite{Vikman04}. When adding a high derivative term to the
kinetic term $X$ in the single-scalar-field model, the
energy-momentum tensor is proven to be equivalent to that of a
two-field quintom model \cite{Limz05}.

In addition, it is remarkable that the generalized ghost condensate
model of a single real scalar field is a successful realization of
the quintomlike dark energy \cite{Tsujikawa:2005ju,holoreczx2}. In
Ref.~\cite{ghost2}, a dark energy model with a ghost scalar field
has been explored in the context of the runaway dilaton scenario in
low-energy effective string theory. The authors addressed for the
dilatonic ghost condensate model the problem of vacuum stability by
implementing higher-order derivative terms and showed that a
cosmological model of quintomlike dark energy can be constructed
without violating the stability of quantum fluctuations.
Furthermore, a generalized ghost condensate model was investigated
in Refs.~\cite{Tsujikawa:2005ju,holoreczx2} by means of the
cosmological reconstruction program. In what follows we will focus
on the generalized ghost condensate model. We shall use this scalar
field model to effectively describe the holographic Ricci dark
energy, and perform the reconstruction for such a scalar model. For
the reconstruction of dark energy models, see, e.g.,
\cite{holoreczx1,holoreczx2,Tsujikawa:2005ju,Saini:1999ba,simplescalar,
Guo:2005at,scalartensor,frgrav,viscosity,Zhang:2006em}.

First, let us consider the Lagrangian density of a general scalar
field $p(\phi, X)$, where
$X=-g^{\mu\nu}\partial_\mu\phi\partial_\nu\phi/2$ is the kinetic
energy term. Note that $p(\phi, X)$ is a general function of $\phi$
and $X$, and we have used a sign notation $(-, +, +, +)$.
Identifying the energy-momentum tensor of the scalar field with that
of a perfect fluid, we can easily derive the energy density
$\rho_{\rm de}=2Xp_X-p$, where $p_X=\partial p/\partial X$. Thus, in
a spatially flat FRW universe, the dynamic equations for the scalar
field are
\begin{equation}
3H^2=\rho_{\rm m}+2Xp_X-p,\label{hsqr}
\end{equation}
\begin{equation}
2\dot{H}=-\rho_{\rm m}-2Xp_X,\label{hdot}
\end{equation}
where $X=\dot{\phi}^2/2$ in the cosmological context. Here we have
used the unit $M_{\rm Pl}=1$ for convenience. Also, for convenience,
we introduce the quantity $r=E^2=H^2/H_0^2$. We find from Eqs.
(\ref{hsqr}) and (\ref{hdot}) that
\begin{equation}
p=[(1+z)r'-3r]H_0^2,\label{p}
\end{equation}
\begin{equation}
\phi'^2p_X={r'-3\Omega_{\rm m0}(1+z)^2\over r(1+z)},\label{px}
\end{equation}
where the prime denotes a derivative with respect to $z$. The
equation of state for dark energy is given by
\begin{equation}
w={p\over \dot{\phi}^2 p_X-p}={(1+z)r'-3r\over 3r-3\Omega_{\rm
m0}(1+z)^3}.
\end{equation}
Next, if we establish a correspondence between the holographic Ricci
vacuum energy and the scalar field dark energy, we should choose a
scalar-field model in which crossing the cosmological-constant
boundary is possible. So, let us consider the generalized ghost
condensate model proposed in Ref.~\cite{Tsujikawa:2005ju}, with the
Lagrangian density
\begin{equation}
p=-X+h(\phi)X^2,
\end{equation}
where $h(\phi)$ is a function in terms of $\phi$. The dilatonic
ghost condensate model \cite{ghost2} corresponds to a choice
$h(\phi)=ce^{\lambda\phi}$. From Eqs.~(\ref{p}) and (\ref{px}) we
obtain
\begin{equation}
\phi'^2={12r-3(1+z)r'-3\Omega_{\rm m0}(1+z)^3\over r(1+z)^2},
\end{equation}
\begin{equation}
h(\phi)={6(2(1+z)r'-6r+r(1+z)^2\phi'^2)\over
r^2(1+z)^4\phi'^4}\rho_{\rm c0}^{-1},
\end{equation}
where $\rho_{\rm c0}=3H_0^2$ represents the present critical density
of the universe. The generalized ghost condensate describes the
holographic Ricci vacuum energy, provided that
\begin{equation}
r(z)={2\over 2-\alpha}\Omega_{\rm m0}(1+z)^3+f_0(1+z)^{(4-{2\over
\alpha})},
\end{equation}
where $f_0=1-2\Omega_{\rm m0}/(2-\alpha)$.

\begin{figure}[htbp]
\begin{center}
\includegraphics[scale=0.5]{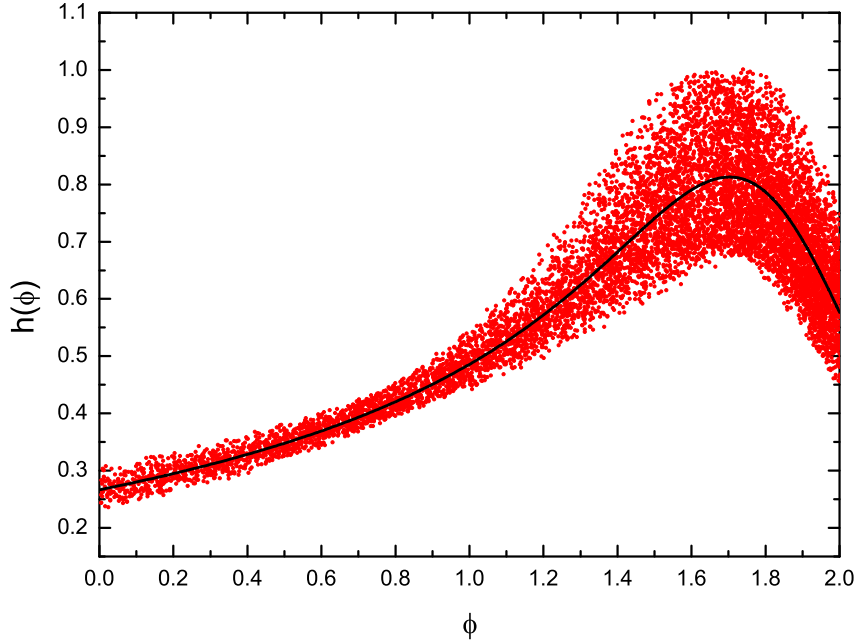}
\caption[]{\small Reconstruction of the generalized ghost condensate
model according to the holographic Ricci dark energy scenario. In
this plot, we show the reconstructed function $h(\phi)$, in units of
$\rho_{\rm c0}^{-1}$, corresponding to the joint analysis results of
SNIa, CMB, and BAO observations. The central black solid line
represents the best fit, and the red dotted area covers the range of
$68\%$ errors. The errors are calculated using a Monte Carlo
method.}\label{fig:hphi}
\end{center}
\end{figure}

\begin{figure}[htbp]
\begin{center}
\includegraphics[scale=1.2]{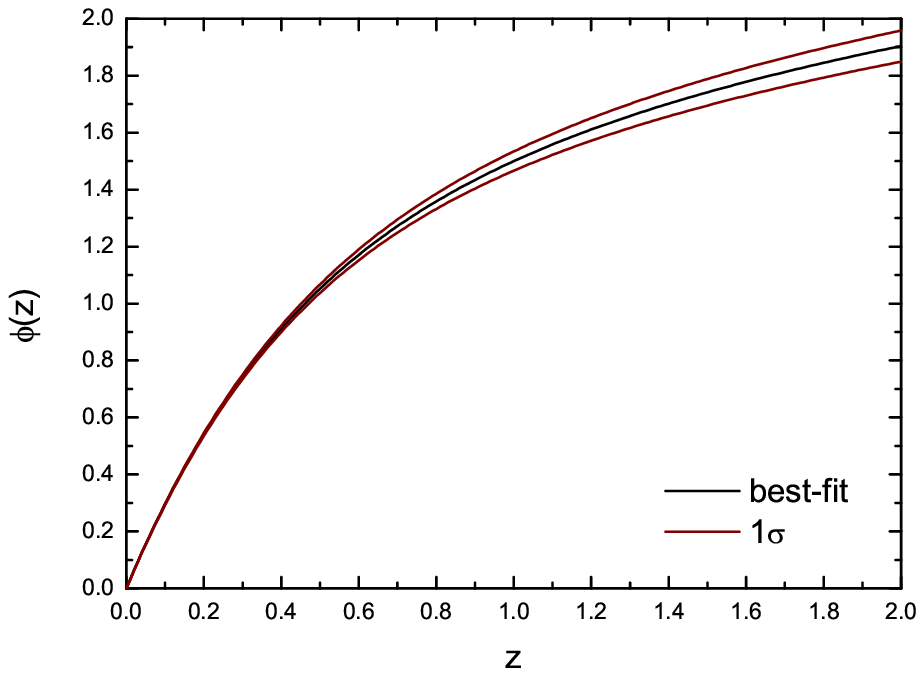}
\caption[]{\small Reconstruction of the generalized ghost condensate
model according to the holographic Ricci dark energy scenario. In
this plot, we show the evolution of the scalar field $\phi(z)$, in
units of the Planck mass $M_{\rm Pl}$ (note that here the Planck
normalization $M_{\rm Pl}=1$ has been used), corresponding to the
joint analysis results of SNIa, CMB, and BAO
observations.}\label{fig:phiz}
\end{center}
\end{figure}

\begin{figure}[htbp]
\begin{center}
\includegraphics[scale=1.2]{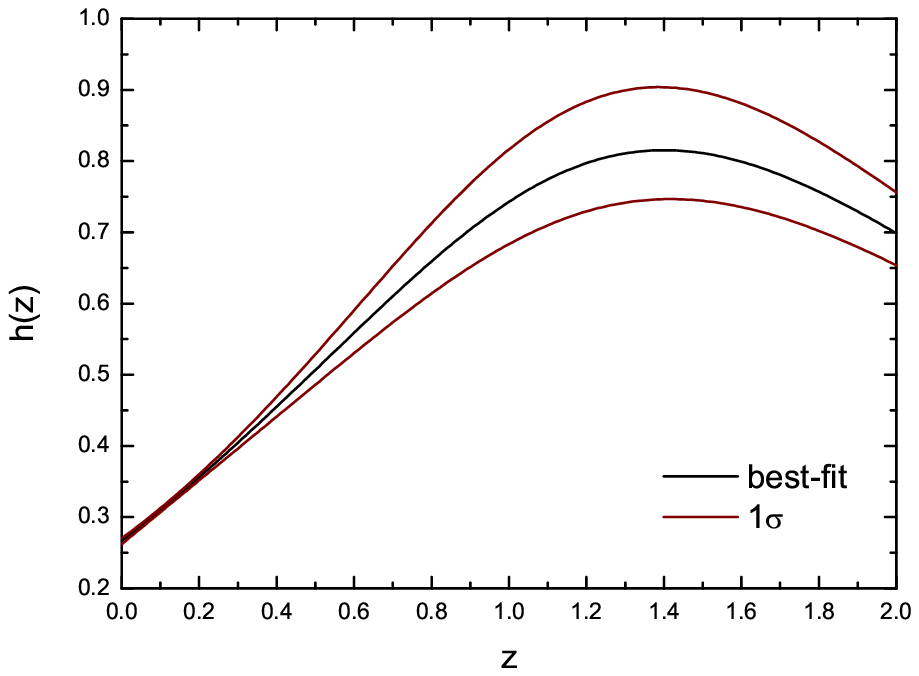}
\caption[]{\small Reconstruction of the generalized ghost condensate
model according to the holographic Ricci dark energy scenario. In
this plot, we show the evolution of the function $h(z)$, in units of
$\rho_{\rm c0}^{-1}$, corresponding to the joint analysis results of
SNIa, CMB, and BAO observations.}\label{fig:hz}
\end{center}
\end{figure}

In Sec.~\ref{sec:obs}, we have derived cosmological constraints on
the holographic Ricci dark energy model from the joint analysis of
SNIa, CMB, and BAO observations. Now, one can reconstruct the
function $h(\phi)$ for the generalized ghost condensate model in
light of the holographic Ricci dark energy and the corresponding fit
results of the observational constraints. The reconstruction for
$h(\phi)$ is plotted in Fig.~\ref{fig:hphi}, using the 1$\sigma$ fit
results from the joint analysis of SNIa, CMB, and BAO observations.
In this figure, the central black solid line represents the best
fit, and the red dotted area around the best fit covers the range of
1$\sigma$ errors. The errors quoted in Fig.~\ref{fig:hphi} are
calculated using a Monte Carlo method where random points are chosen
in the 1$\sigma$ region of the parameter space shown in
Fig.~\ref{fig:contour}. The evolution of the scalar field $\phi(z)$
is also determined by the reconstruction program (see
Fig.~\ref{fig:phiz}) in which we have fixed the field amplitude at
the present epoch ($z=0$) to be zero: $\phi(0)=0$. In addition, the
reconstructed evolution of $h(z)$ is also shown in
Fig.~\ref{fig:hz}. Note that the errors quoted in
Figs.~\ref{fig:phiz} and \ref{fig:hz} are calculated using the
Fisher matrix approach.

The crossing of the cosmological-constant boundary corresponds to
$hX=1/2$. The system can enter the phantom region ($hX <1/2$)
without discontinuous behavior of $h$ and $X$. In addition, as has
been pointed out by Tsujikawa \cite{Tsujikawa:2005ju}, it should be
cautioned that the perturbation of the field $\phi$ is plagued by a
quantum instability whenever it behaves as a phantom \cite{ghost2}.
Even at the classical level, the perturbation becomes unstable for
$1/6<hX<1/2$, because the speed of sound $c_s^2=p_X/(p_X+2Xp_{XX})$
will become negative. This instability may be avoided if the phantom
behavior is just transient. In fact the dilatonic ghost condensate
model can realize a transient phantom behavior (see, e.g., Fig.~4 in
Ref.~\cite{ghost2}). In this case the cosmological-constant boundary
crossing occurs again in the future, after which the perturbation
will become stable. Nevertheless, one may argue that the field can
be regarded as an effective one so as to evade problems such as
stability. In particular, the present focus is how to establish a
dynamical scalar-field model on a phenomenological level to describe
the possible underlying theory of dark energy, disregarding whether
the field is fundamental or not.

\section{Concluding remarks}\label{sec:concl}

The cosmic acceleration observed by distance-redshift relation
measurement of SNIa strongly supports the existence of dark energy.
The fantastic physical property of dark energy not only drives the
current cosmic acceleration, but also determines the ultimate fate
of the universe. However, hitherto, the nature of dark energy as
well as its cosmological origin still remain enigmatic for us.
Though the underlying theory of dark energy is still far beyond our
knowledge, it is guessed that the quantum gravity theory shall play
a significant role in resolving the dark energy enigma.

Therefore, one can try to probe the nature of dark energy according
to some principles of quantum gravity. By far, the holographic
principle is widely believed as a fundamental principle for the
theory of quantum gravity. So, the holographic models of dark energy
become an important attempt for exploring dark energy within the
framework of quantum gravity. It is believed that the holographic
models possess some significant features of an underlying theory of
dark energy.

In this paper, we consider the model of holographic Ricci dark
energy that can be viewed as originating from taking the causal
connection scale as the IR cutoff in the holographic setting. The
main characteristic of holographic Ricci dark energy is governed by
a positive numerical parameter $\alpha$ in the model. In particular,
when $\alpha<1/2$, the holographic Ricci dark energy will exhibit a
quintomlike behavior; i.e., its equation of state will evolve across
the cosmological-constant boundary $w=-1$. The parameter $\alpha$
can be determined only by observations. Thus, in order to
characterize the evolving feature of dark energy and to predict the
fate of the universe, it is of extraordinary importance to constrain
the parameter $\alpha$ by using the observational data.

We have derived, in this paper, the constraints on the holographic
Ricci dark energy model from the latest observational data including
the 307 Union sample of SNIa, the CMB shift parameter given by
WMAP5, and the BAO measurement from SDSS. The joint analysis gives
the best-fit results (with 1$\sigma$ confidence level):
$\alpha=0.359^{+0.024}_{-0.025}$ and $\Omega_{\rm
m0}=0.318^{+0.026}_{-0.024}$. That is to say, according to the
observations, the holographic Ricci dark energy takes on a quintom
feature.

If we regard the scalar-field model as an effective description of
such a theory (holographic Ricci vacuum energy), we should be
capable of using the scalar-field model to mimic the evolving
behavior of the dynamical vacuum energy and reconstructing this
scalar-field model according to the evolutionary behavior of
holographic Ricci dark energy and the fits to the observational data
sets. We find the generalized ghost condensate model is a good
choice for depicting the holographic Ricci vacuum energy, since it
can easily realize the quintom behavior. We thus reconstructed the
function $h(\phi)$ of the generalized ghost condensate model using
the fit results of the observational data (SNIa + CMB + BAO). We
hope that the future high precision observations (e.g., the
SuperNova Acceleration Probe) may be capable of determining the fine
property of the dark energy and consequently reveal some significant
features of the underlying theory of dark energy.

\section*{Acknowledgements}
The author thanks Chang-Jun Gao, Yan Gong, You-Gang Wang, Hao Wei,
and Feng-Quan Wu for useful discussions. Special thanks go to
Xiao-Dong Li for kind help in the program. This work was supported
by the National Natural Science Foundation of China (Grant No.
10705041).


\end{document}